# Hanny's Voorwerp: a nuclear starburst in IC2497

### M. A. Garrett[1-3]
[1]*ASTRON, Netherlands Institute for Radio Astronomy, Post box 2, 7990AA, Dwingeloo, The Netherlands.*
[2]*Leiden Observatory, Leiden University, Post box 9513, 2300RA Leiden, The Netherlands.*
[3]*Centre for Astrophysics and Supercomputing, Swinburne University of Technology, Australia.*
*E-mail:* `garrett@astron.nl`

### H. Rampadarath
*Curtin University of Technology, GPO BOX U1987, Perth, WA 6845, Australia.*
*E-mail:* `Hayden.Rampadarath@icrar.org`

### G. I. G. Józsa
*ASTRON, Netherlands Institute for Radio Astronomy, Postbus 2, 7990AA, Dwingeloo, The Netherlands.*
*E-mail:* `jozsa@astron.nl`

### T. W. B. Muxlow
*Jodrell Bank Centre for Astrophysics, School of Physics and Astronomy, Univ. Manchester, Alan Turing Building, Oxford Road, Manchester, M13 9PL, UK.*
*E-mail:* `Tom.Muxlow@manchester.ac.uk`

### T. A. Oosterloo
*ASTRON, Netherlands Institute for Radio Astronomy, Postbus 2, 7990AA, Dwingeloo, The Netherlands.*
*Kapteyn Astronomical Institute, Univ. Groningen, Postbus 800, 9700 AV Groningen, The Netherlands.*
*E-mail:* `oosterloo@astron.nl`

### Z. Paragi
*JIVE, Joint Institute for VLBI in Europe, Postbus 2, 7990AA, Dwingeloo, The Netherlands.*
*MTA Research Group for Physical Geodesy and Geodynamics, PO Box 91, 1521 Budapest, Hungary.*
*E-mail:* `paragi@jive.nl`

### R. Beswick
*Jodrell Bank Centre for Astrophysics, School of Physics and Astronomy, Univ. Manchester, Alan Turing Building, Oxford Road, Manchester, M13 9PL, UK.*
*E-mail:* `Robert.Besick@manchester.ac.uk`

**H. van Arkel**

*ASTRON, Netherlands Institute for Radio Astronomy, Postbus 2, 7990AA, Dwingeloo, The Netherlands.*
*E-mail:* `hanny.van.arkel@home.nl`

**K. Schawinski**

*Univ. Yale, Dept. Physics, J.W. Gibbs Laboratory, 260 Whitney Avenue, Yale University, New Haven, CT 06511, USA.*
*E-mail:* `kevin.schawinski@yale.edu`

**W. C. Keel**

*Univ. Alabama, Dept. Physics & Astronomy, Box 870324, University of Alabama, Tuscaloosa, AL 35487-0324, USA.*
*E-mail:* `wkeel@bama.ua.edu`

We present high and intermediate resolution radio observations of the central region in the spiral galaxy IC 2497, performed using the European VLBI Network (EVN) at 18 cm, and the Multi-Element Radio Linked Interferometer Network (MERLIN) at 18 cm and 6 cm. The e-VLBI observations detect two compact radio sources with brightness temperatures in excess of $10^5$ K, suggesting that they are associated with an AGN located at the centre of the galaxy. We show that IC2497 lies on the FIR-radio correlation and that the dominant component of the 18 cm radio flux density of the galaxy is associated with extended emission confined to sub-kpc scales. IC 2497 therefore appears to be a luminous infrared galaxy that exhibits a nuclear starburst with a total star formation rate (assuming a Salpeter IMF) of ~ 70 $M_\odot$/yr. Typically, vigorous star forming galaxies like IC2497 always show high levels of extinction towards their nuclear regions. The new results are in-line with the hypothesis that the ionisation nebula "Hanny's Voorwerp", located ~15−25 kpc from the galaxy is part of a massive gas reservoir that is ionised by the radiation cone of an AGN that is otherwise obscured along the observer's line-of-sight.

## 1. Introduction

Hanny's Voorwerp (SDSS J094103.80+344334.2) is an irregular gas cloud located ∼ 25 kpc to the southeast of the massive disk galaxy IC 2497 [1]. In the optical, the Voorwerp's appearance is dominated by [O III] emission lines, and its spectrum shows strong line emission, with high-ionisation lines co-extensive with the continuum. Paradoxically, there is no clear evidence of an ionising source in the immediate proximity of this nebulosity.

The original interpretation of this phenomenon [1], suggested that Hanny's Voorwerp may be the first example of a quasar light echo. Indeed, Lintott et al. proposed that in the past, the central luminosity of IC2497 approached quasar-like levels but that around $10^5$ years ago this decreased to the lower-levels of activity observed today. X-ray observations [2] are also interpreted in this context – there it is concluded that the galaxy's central engine has decreased its radiative output by at least 2 orders of magnitude within the last 70,000 years. IC 2497 is a quasar that has transitioned from a high state to a radiatively inefficient state where the bulk of the energy is dissipated not as radiation but as either thermal or kinetic energy. Radio observations using the Westerbork Synthesis Telescope (WSRT) and the European VLBI Network (EVN) have detected a radio continuum source at the central position of IC 2497, and weak, one-sided, large-scale emission (interpreted as a radio jet) pointing exactly in the direction of Hanny's Voorwerp [3]. In addition, neutral hydrogen, presumably debris from a past interaction, is detected around the galaxy. The radio observations strongly suggest that the Voorwerp is part of a huge reservoir of gas that surrounds IC2497. While HI is detected in emission, it is also detected in absorption towards the central radio core. These radio observations (see Fig.1) lead to an alternative interpretation of the Voorwerp, namely that IC 2497 contains an obscured active galactic nucleus (AGN) with a weak, large-scale radio jet pointing in the direction of Hanny's Voorwerp, perpendicular to the major axis of the galaxy [3].

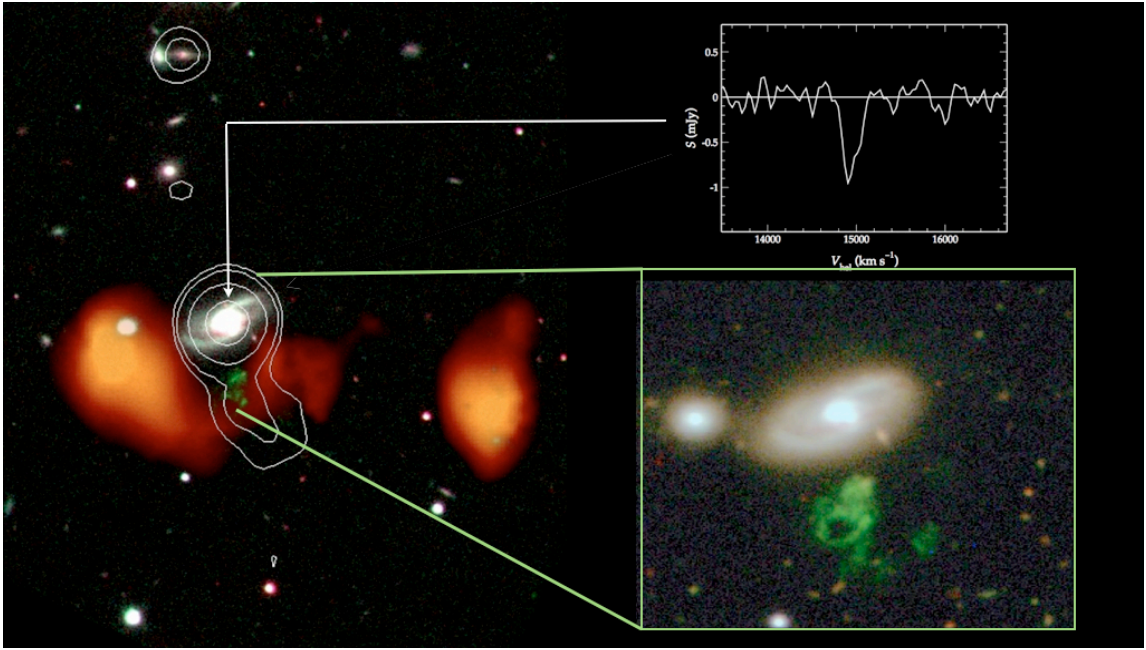

*Fig.1 Summary slide of the WSRT results – the HI gas observed in emission around IC2497 is displayed in orange, the white contours represent the radio continuum emission and the HI absorption feature is shown top right (Józsa et al. 2009). IC2497 and Hanny's Voorwerp are shown bottom right (courtesy W. Keel, D. Smith, P. Herbert, and M. Jarvis).*

We present new radio observations of IC2497 made with the EVN using the e-VLBI technique at 18 cm, and with the Multi-Element Radio Linked Interferometer Network (MERLIN) at 18 cm and 6 cm. The observations confirm the earlier detection of compact radio sources in the central region of IC2497 [3] but also reveal that the core of the galaxy hosts a luminous nuclear starburst system.

## 2. EVN (e-VLBI) and MERLIN maps of IC2497

On 19-20 May 2009, IC 2497 was observed at 18 cm with the EVN using the e-VLBI technique. The total on-source integration time on the target was ~ 7 hours. Additional observations of IC 2497 were conducted with MERLIN at 18 (2-3 February 2009 and 4-5 February 2009 with a total on-source integration time of ~ 27 h) and 6 cm (21-23 March 2009 with a total on-source integration time of 32 h). Both the EVN and MERLIN observations employed standard phase-referencing techniques. The naturally weighted EVN and MERLIN 18cm images [4] are shown together in Fig.2. The 18cm EVN image shows that two distinct and compact radio sources are detected in the central region of IC 2497 (labeled C1 and C2). The MERLIN 18cm image shows an extended region of emission that is confined to sub-kpc scales. The MERLIN 6cm image (see [4]) detects only 1 compact source, C2.

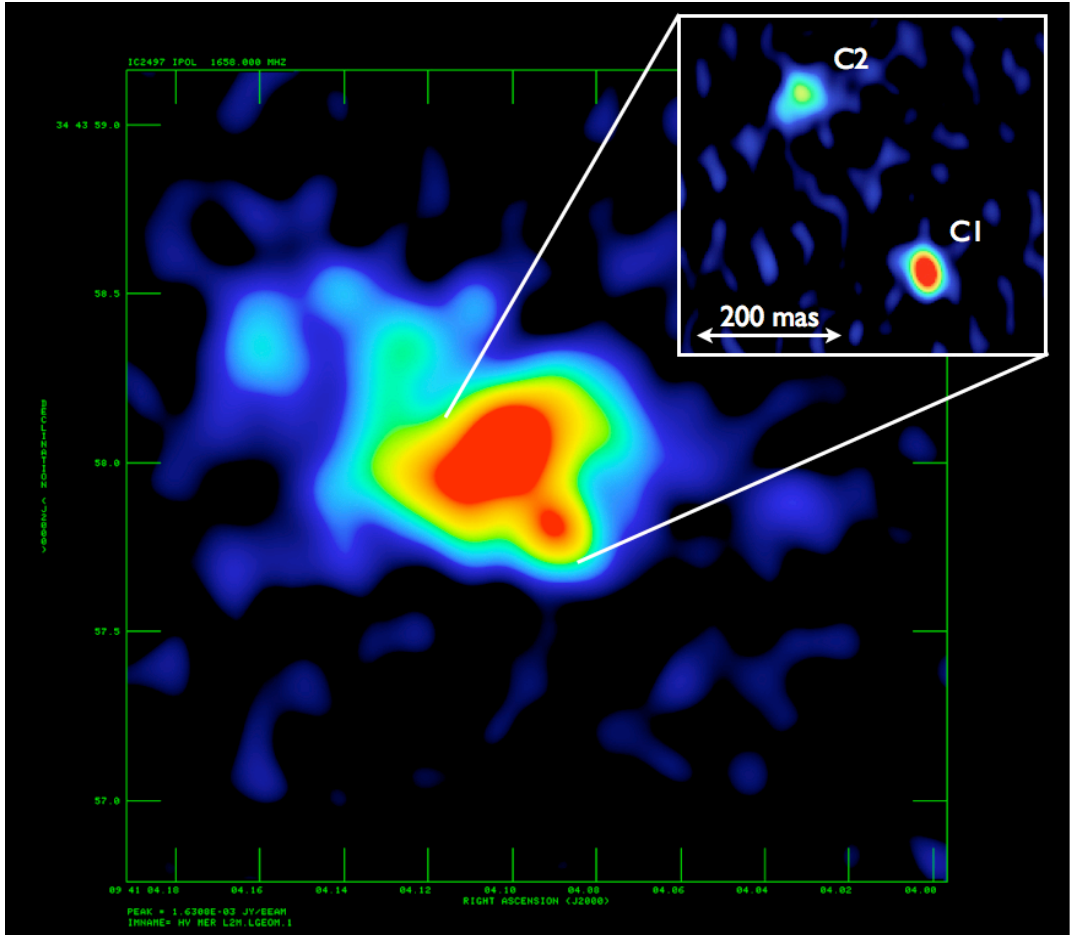

*Figure 2: The MERLN 18cm image and VLBI 18cm image (inset).*

## 3. Evidence for AGN activity in IC2497

The brightness temperatures of both the radio components detected by the EVN at 18cm is in excess of $10^5$ K, an upper limit for the brightness temperature of star-forming regions (e.g. [5]). Comparing the appropriately tapered EVN 18cm image with the MERLIN 6 cm image, we derive a relatively flat spectral index ($S \propto \nu^{\alpha}$) for $\alpha_{C2} \sim 0.12 \pm 0.01$, suggesting that this is the radio core that corresponds to the central AGN in IC 2497. By comparison $C_1$, has a much steeper spectrum: assuming an upper limit to the MERLIN flux density of $C_1 \sim 0.26$ mJy, we derive a spectral index of $\alpha_{C1} < -1.38 \pm 0.10$. In this scenario, $C_1$ is most likely a hotspot in the large-scale jet observed on larger scales. Unfortunately, the radio luminosity of $C_2$ tells us very little about the ionising potential of the associated AGN at optical and UV wavelengths. It does clearly indicate however, some level of AGN activity in IC 2497. The approximate separation of the components $C_1$ and $C_2$ is 300 milliarcseconds, with the associated position angle being 215 degrees. Given the different scales involved, this is very similar to the position angle defined by both the direction of the WSRT kpc jet [3] and the Voorwerp itself of ~ 280 degrees. Our results support the hypothesis that an AGN is located at the centre of IC 2497.

## 4. Evidence for a nuclear starburst in IC2497

The extended emission clearly detected by MERLIN at 18 cm and associated with $C_2$ is aligned with the major axis of the galaxy. This extended emission has physical dimensions of ~ 0.4 kpc. The MERLIN 18cm image recovers ~ 75% of the emission observed by the VLA FIRST survey and ~ 70% of the emission observed by the WSRT in the central, unresolved component [3]. This suggests that most of the radio emission measured by the WSRT and VLA is also associated with the extended emission detected in the MERLIN image. We estimate a flux density for the extended radio emission of 15.7 ± 1.1 mJy. This, in addition to the IRAS FIR flux densities of $S_{100\mu} = 3.04 \pm 0.3$ Jy and $S_{60\mu} = 1.66 \pm 0.2$ Jy [6] permits us to calculate a q-value of 2.2 ± 0.04. Clearly, IC 2497 lies close to the standard FIR-radio correlation for star-forming galaxies, and the sub-kpc scale of the extended emission implies that it is very likely a nuclear starburst system. It appears that this nuclear starburst is coincident with the AGN core. Using the standard relations given in [5] and assuming a Salpeter IMF, we derive a total star formation rate ~ 70 $M_\odot$ yr$^{-1}$ for the nuclear region.

## 5. Hanny's Voorwerp – the radio view

Radio observations with the WSRT, VLA, MERLIN and e-VLBI provide an important and indeed unique view of Hanny's Voorwerp –one that is free of the effects of dust obscuration. The WSRT observations of HI in emission demonstrate that Hanny's Voorwerp is part of a massive reservoir of cold gas that surrounds IC2497. The origin of the gas is probably due to past interactions between IC2497 and other nearby systems. The radio continuum (on all scales) suggests that an active AGN lies at the heart of IC2497 and that a jet is present. This jet is perpendicular to the disk of IC2497 and appears to point in the direction of the Voorwerp. However, extended radio emission on sub-kpc scales, together with previous infrared measurements, strongly suggest that the bulk of the radio emission in IC2497 is associated with a nuclear starburst. This appears to be coincident with the central AGN. In the nearby Universe, galaxies like IC2497 often present co-eval AGN and nuclear star formation activity fueled by past or on-going interactions. Typically, galaxies with total star formation rates of ~ 10-100 $M_\odot$/yr also present evidence of significant levels of extinction due to dust – Arp 220 and M82 are good examples of this. The earlier detection by the WSRT of HI absorption towards IC2497 already provides direct evidence for obscuration towards the nuclear regions.

Fig. 2 shows a sketch of the hypothesis that best explains the radio observations presented here. In this scenario, the line-of-sight between IC2497 and the observer is obscured by the nuclear starburst that lies in the same plane as the larger-scale galactic disk. The line-of-sight perpendicular to the disk is largely un-obscured – due to the favourable orientation of the disk and possibly via the action of the jet disrupting the gas and dust of the ISM along this direction. A cone of hard radiation therefore emerges from the central AGN and illuminates a limited segment of the massive gas reservoir that resides in the vicinity of IC2497. This illuminated segment is therefore ionised by the direct radiation from the AGN, and presents the phenomena that we know as "Hanny's Voorwerp".

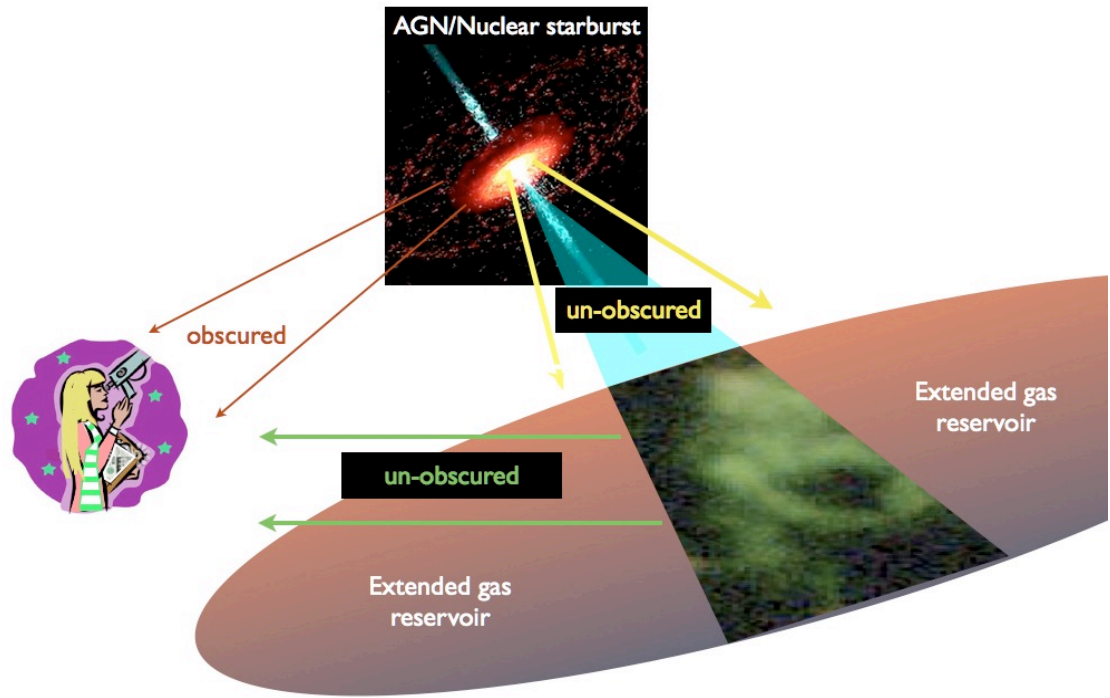

*Fig.2 A sketch of the scenario described in the text that best explains the radio observations.*